\documentclass[aps,prl,twocolumn,showpacs]{revtex4}
\usepackage{graphicx,epsfig}
\begin{document}
%\draft
\title{Temperature dependent magnetic
anisotropy in metallic magnets from an ab-initio electronic structure
theory: $L1_0$-ordered $FePt$} 

\author{J.B.Staunton$^1$, S.Ostanin$^{2}$, S.S.A.Razee $^{3}$,
B.L.Gyorffy $^{4}$, L.Szunyogh $^{5}$, B.Ginatempo$^{6}$ and 
Ezio Bruno$^{6}$}
 \affiliation{$^1$Department of Physics, University of Warwick, Coventry CV4 7AL,
U.K.}

\affiliation{$^2$ Department of Earth Sciences, University College London,
Gower Street, London WC1E 6BT, U.K. }

\affiliation {$^3$Department of Physics, Kuwait University, P.O. Box 5969,
SAFAT 13060, Kuwait }

\affiliation{$^4$ H.H.Wills Physics Laboratory, University of Bristol, Tyndall 
Avenue, Bristol BS8 1TL, U.K.}

\affiliation{$^5$ Department of Theoretical Physics, Budapest 
University of Technology and Economics, Budapest, Hungary}

\affiliation{$^6$ Dipartimento di Fisica and Unita INFM, Universita di Messina,
Sicily, Italy}

\date{\today}
%\maketitle

\begin{abstract}

On the basis of a first-principles, relativistic electronic structure theory
of finite temperature metallic magnetism, we investigate the variation of 
magnetic anisotropy, $K$, with magnetisation, $M$, in metallic ferromagnets.
We apply the theory to the high magnetic anisotropy
material, $L1_0$-ordered FePt, and find its uniaxial $K$ consistent with a 
magnetic easy axis perpendicular to the Fe/Pt layering for all $M$ and 
to be proportional to $M^2$ for a broad range of values of $M$.
For small $M$, near the Curie temperature, the calculations pick out the 
easy axis for the onset of magnetic order. Our results 
are in good agreement with recent experimental measurements on this important 
magnetic material.  
\end{abstract}
\pacs{75.30.Gw, 75.10.Lp, 71.15.Rf,75.50.Bb,75.50.Ss}
\maketitle

By accounting for relativistic effects such as spin-orbit coupling
on electronic structure, recent `first-principles' theoretical work has 
been successful in describing trends in the magnetocrystalline anisotropy (MCA)
of magnetic materials.~\cite{Jansen,Razee+99,refs} This is useful for the 
understanding of permanent magnetic properties, domain wall structure, magnetic 
nanostructures etc. One aspect, however, which has received scant attention 
from such ab-initio theories is its temperature dependence.  
Modelling this phenomenon seems to rest largely on the seminal papers
by Callen and Callen and others~\cite{Callen} published nearly 40 years ago
which focussed on the temperature dependence associated with single ion
magnetic anisotropies. In this letter we investigate how far this approach
can be justified for metallic magnets on the basis of a 
first-principles, material specific, parameter-free theory of how
the magnetocrystalline anisotropy depends on temperature. The
proposed theory involves a detailed, relativistic description of
the electronic structure and hence it includes a complete 
desciption of the spin-orbit coupling. The thermally induced 
magnetic fluctuations are taken into account by a relativistic 
generalisation of the, by now well-tried, 
disordered local moment (DLM) picture.~\cite{Moriya,DLM}

The topic has recently received extra impetus from extensive experimental
studies of magnetic films and nanostructures and 
their technological potential. For example, 
the recent fabrication of assemblies of smaller and smaller magnetic
nanoparticles holds considerable promise for the design of ultra-high
density magnetic data storage media.~\cite{Sun}  But this is hampered by
a particle size limit set so that
thermally driven demagnetisation and loss of data is avoided
over a reasonable storage period. This limit can be lowered by using 
materials with high magnetocrystalline anisotropy, $K$, since the 
superparamagnetic diameter of a magnetic particle is proportional to $(k_B T/K)
^{\frac{1}{3}}$, where $k_B T$ is the thermal energy.~\cite{OHandley}
In this context,
the chemically ordered $L1_0$ phase of equiatomic $FePt$, which has high
uniaxial MCA (4-10 10$^7$ ergs/cm$^3$ or up to 1.76 meV per $FePt$ pair
~\cite{FePt-MAE,Farrow}), has been 
subjected to much attention and arrays of $FePt$ nanoparticles with
diameters as small as 3 nm have been synthesised.~\cite{Sun,Wu}
A way to write to media of very high $K$ material is by temporary 
heating.~\cite{Thiele2002,Lyberatos,TAR} The 
magnetocrystalline anisotropy
is reduced significantly during the magnetic write process and the 
information is locked in as the material cools. 
Modelling this process and improving the design of high density 
magnetic recording media therefore requires an understanding of how $K$ 
varies with temperature. 

So for the first application of our theory we have chosen 
$L1_0$-ordered $FePt$. 
Given its technological potential there have 
recently been some careful experimental studies of its fundamental magnetic 
properties.~\cite{Okamoto,Thiele2002,Wu} These show a strong 
temperature dependence to $K$. We find good agreement with these
data. In particular we find $K(T) \propto (M(T)/M(0))^2$  
over a broad magnetisation range, in line with the results reported in 
two of these three studies.~\cite{Okamoto,Thiele2002} Notably the low 
temperature behavior 
is qualitatively different from that of the single-ion anisotropy
models used over many years.~\cite{Callen}
 
The magnetic anisotropy of a material can be conveniently expressed as
$K=\sum_{\gamma} k_{\gamma} g_{\gamma} ( \hat{n})$ where the $k_{\gamma}$'s are magnetic 
anisotropy constants, $\hat{n}$ is the direction of
the magnetisation and $g_{\gamma}$'s are polynomials (spherical harmonics) of the 
angles $\theta$, $\phi$ describing the orientation of  
$\hat{n}$ and belong to the fully symmetric representation of the 
crystal point group. For a uniaxial ferromagnet such as $L1_0$-ordered 
$FePt$, $K= k_0 + k_2 ( \cos^2 \theta - 1/3) + \cdots$. 

As the temperature rises, $K$ decreases rapidly. The key features
of the results of the early theoretical work on this 
effect~\cite{Callen} are revealed by a classical spin model pertinent to 
magnets with localised magnetic moments. The anisotropic behavior of 
a set of localised `spins' associated
with ions sitting on crystalline sites,$i$, in the material, is given by
a term in
the hamiltonian $H_{an} = \sum_i \sum_{\gamma} k_{\gamma} g_{\gamma} ( \hat{s}_i)$ with 
$\hat{s}_i$ a unit vector denoting the spin direction on the site $i$. 
As the
temperature is raised, the `spins' sample the energy surface over a small
angular range about the magnetisation direction and the anisotropy energy
is given from the difference between averages taken for the magnetisation 
along the easy and hard directions. If the coefficients $k_{\gamma}$ are assumed 
to be rather insensitive to temperature, the dominant thermal variation 
of $K$ for a ferromagnet is given by 
$K(T)/K(0) = \langle g_{l} ( \hat{s})
\rangle_{T}/\langle g_{l} ( \hat{s} ) \rangle_{0}$
The averages $\langle \cdots \rangle_{T}$ are taken such that $\langle \hat{s}
\rangle_{T} = M(T)$, the magnetisation of the system at temperature $T$, and
 $l$ is the order of the spherical harmonic
describing the angular dependence of the local anisotropy i.e. $l =2$ and
$4$ for uniaxial and cubic systems respectively. At low temperatures 
$K(T)/K(0) \approx (M(T)/M(0))^{l(l+1)/2}$ and near the Curie 
temperature $T_c$, $K(T)/K(0) \approx (M(T)/M(0))^{l}$. These features
are borne out rather well in magnets where the magnetic moments are 
well-localised, e.g. rare-earth and oxide magnets, but it is
questionable whether such an analysis should hold for itinerant 
ferromagnets.~\cite{OHandley}
Here, we examine these issues in the context of the 
$L1_0$-ordered FePt intermetallic compound for which 
careful experiments~\cite{Okamoto,Thiele2002} find  
$K(T)/K(0)=(M(T)/M(0))^n$ ,where $n=2$ instead of $n=3$, over a
large temperature range. As will be shown presently, our 
ab-initio calculations are in good agreement with this 
surprising result.

Magnetocrystalline anisotropy is caused largely by spin-orbit coupling and 
receives an ab-initio description from the relativistic generalisation of  
spin density functional (SDF) theory.~\cite{Jansen} Up to now calculations of 
the anisotropy constants $K$ have been suited to $T=0$K only. They treat
the spin-orbit coupling effects using either perturbation 
theory~\cite{refs} or a fully relativistic one~\cite{Razee+97,Razee+99}. 
Typically the total energy, 
or the single-electron contribution to it (if the force theorem
is used), is calculated for two or more magnetisation directions, 
$\hat{n}_1$ and $\hat{n}_2$ separately, i.e. $E_{\hat{n}_1}$, 
$E_{\hat{n}_2}$, and then the MCA is obtained from the difference, $\Delta E$.
Since the MCA is a small part of the 
total energy of the system, in many cases of the order of $\mu eV$,
it is numerically more precise to calculate the difference directly 
\cite{Razee+97} so that $\Delta E = - \int^{E_{F_1}} [ N ( 
\varepsilon;\hat{n}_1 ) - N (\varepsilon;\hat{n}_2 ) ] d \varepsilon
- \frac{1}{2} n ( E_{F_2};\hat{n}_2 ) ( E_{F_1} - E_{F_2} )^2 +
{\mathcal O} ( E_{F_1} - E_{F_2} )^3$ where $E_{F}^{1}$, $E_{F}^{2}$ 
are the Fermi energies when the
system is magnetised along the directions $\hat{n}_{1}$ and $\hat{n}_{2}$
respectively and $n(\varepsilon;\hat{n})$ and $N(\varepsilon;\hat{n})$ 
are the density of states and integrated density of states respectively. 
We have used this rationale with a fully relativistic theory to study the 
MCA of magnetically soft, compositionally disordered binary and ternary 
component alloys~\cite{Razee+97,more}, the effect upon it of atomic 
short-range order~\cite{Razee+99} and recently its dependence in $FePt$ 
upon the extent of $L1_0$ long range chemical order~\cite{ICNDR}. 

There are a number of calculated values of the MCA 
of completely $L1_0$-ordered $FePt$ at $T=0$K in the 
literature~\cite{FePtMAE-calcs} ranging from 1.2 to 3.9 meV (7-22 10$^7$)
ergs/cm$^3$. $K$ for $FePt$ has been measured at room 
temperature~\cite{FePt-MAE}
to be 6.6 10$^7$ergs/cm$^3$ although more recent measurements~\cite{Farrow}
suggest that this value could exceed 10$^8$ergs/cm$^3$ for completely ordered
$FePt$. The easy axis in both experiment and theory is along the c-axis,
$(0,0,1)$, perpendicular to the $Fe$ and $Pt$ layers. 

Our calculations of the MCA of materials use spin-polarised, relativistic 
multiple
scattering theory and an adaptive mesh algorithm~\cite{EB+BG} for Brillouin 
zone integrations such that the numerical precision of our calculations 
is to within 0.1~$\mu$eV.~\cite{Razee+97,Razee+99} These attributes are also
important for the theoretical calculations of the temperature dependence of 
the MCA described below. 
We calculate the MCA of ordered FePt at $T=0$K to be 1.696 meV. 
We start from a self-consistent field, scalar
relativistic calculation (atomic sphere approximation) of the 
electronic structure and effective potentials for the Fe and Pt sites. We then
perform a further fully relativistic electronic structure calculation,
recalculate the Fermi energies $E_{F}^{1}$ and $E_{F}^{2}$ and determine the  
MCA. For $L1_0$-ordered FePt we find that if $E_{F}^{1}$ and $E_{F}^{2}$
are artificially changed by a small amount, the MCA varies significantly. For
example,the MCA jumps from 1.696 to 2.751 meV if the $E_{F}$'s are lowered 
by just 0.2 eV. This sensitivity may explain, in part, the range of published 
values of the MCA of FePt.
We can also deduce that the magnetic anisotropy of this strong permanent magnet
 might be further enhanced
by replacing a few atomic percent of Pt with Ir. 

In a metallic ferromagnet at $T=0$K the electronic band structure is
spin-polarised. With increasing temperature, spin fluctuations are induced 
which eventually
destroy the long-range magnetic order and hence the overall spin polarization 
of the system's electronic structure. These collective electron modes interact
as the temperature is raised and are dependent upon and affect the underlying
electronic structure. For many materials the magnetic excitations can be
modelled by associating local spin-polarisation axes with all lattice sites and
the orientations $\{ \hat{e}_i \}$ vary very slowly on the time-scale of the
electronic motions.~\cite{Moriya} These 
`local moment' degrees of freedom produce local 
magnetic fields on the lattice sites which affect the electronic motions and
are self-consistently maintained by them. By taking appropriate ensemble 
averages over the orientational configurations, the magnetic 
properties of a system can be determined and, with the explicit inclusion of
relativistic effects upon the electronic structure, the temperature dependence 
of the MCA can be obtained.
      
Consider this DLM picture of a ferromagnetic metal magnetised along a 
direction $\hat{n}$ at a temperature $T$. The
probability that the system's local moments are configured according to  
$\{ \hat{e}_i \}$ 
is
\begin{equation} 
P^{(\hat{n})} ( \{ \hat{e}_i \} ) = \exp [ - \beta \Omega^{(\hat{n})} 
( \{ \hat{e}_i \} ) ]/Z^{(\hat{n})}
\end{equation}
where the partition function  
$Z^{(\hat{n})} =$ $ \prod_j \int d \hat{e}_j \exp [ - \beta \Omega^{(\hat{n})}
 ( \{ \hat{e}_i \} ) ] $.
$ \Omega^{(\hat{n})} ( \{ \hat{e}_i \} ) $  is the `generalised'
electronic grand potential from SDFT~\cite{DLM} and $ \beta = 
(k_B T ) ^{-1}$. The thermodynamic free energy, which accounts for the 
entropy associated with the orientational fluctuations as well as creation 
of electron-hole pairs, is given by $F^{(\hat{n})}= -k_{B} T \log 
Z^{(\hat{n})}$. The role of a local moment
 hamiltonian, albeit a highly complicated one is played by $\Omega \{
  \hat{e}_{i}\}$. By choosing a suitable reference `spin' hamiltonian
$\Omega_{0}\{ \hat{e}_{i}\} = \sum_{i} h^{(\hat{n})} \hat{n} \cdot 
\hat{e}_i$
 and, using the Feynman Inequality~\cite{Feynman}, a mean field
theoretical estimate of the free energy is obtained,~\cite{DLM}
\begin{equation}
F^{(\hat{n})} = \langle \Omega^{(\hat{n})} \rangle + (1/\beta) \sum_i
\int P^{(\hat{n})} (\hat{e}_i) \ln P^{(\hat{n})} (\hat{e}_i) d\hat{e}_i
\end{equation}
where the probability of a moment pointing along $\hat{e}_i$ on a site $i$ is
\begin{equation} 
P^{(\hat{n})} (\hat{e}_i) = \frac{\exp[-\beta h^{(\hat{n})}
\hat{n} \cdot \hat{e}_i] }{ \int \exp[-\beta h^{(\hat{n})} \hat{n}
\cdot \hat{e}_i]  d \hat{e}_i} 
\end{equation}
 and the Weiss field at a site is given by
\begin{equation}
h^{(\hat{n})} =\frac{3}{4 \pi} \int \langle \Omega^{(\hat{n})}
\{ \hat{e}_l \} \rangle_{\hat{e}_i} \hat{n} \cdot \hat{e}_i
d \hat{e}_i.
\end{equation}
where $\langle \cdots \rangle_{\hat{e}_i}$ denotes a constrained
statistical average with the magnetic moment on site $i$ being 
fixed along $\hat{e}_i$.
 
The magnetisation  ${\bf M} =M \hat{n}$ is given by
$M = \mu \int P^{(\hat{n})} (\hat{e}_i)  \hat{n} 
\cdot \hat{e}_i d \hat{e}_i$. $\mu$ is the size of the
local moment on the site and is determined self-consistently.~\cite{DLM}
For materials where this DLM picture is suitable, the sizes of the local 
moments, $\mu$, remain fairly constant so that even in the paramagnetic 
state where $M=0$, the $\mu$'s are roughly the same as the magnetic 
moment per atom in the ferromagnetic state at $T=0$K. 
In a first-principles implemenation of such a DLM picture,
the averaging over the orientational configurations of the
local moments is performed using the KKR-CPA method
adopted from the theory of random metallic 
alloys.~\cite{SCF-KKR-CPA,DLM} Using this methodology over the
past 20 years,
the paramagnetic state, onset of magnetic order and transition 
temperatures of many systems have been successfully 
described.~\cite{DLM-applics} All applications to date, however, have neglected
relativistic effects and have been devoted 
to the paramagnetic state where the 
symmetry turns the calculation into a binary alloy-type one with 
half the moments oriented along a direction and the rest antiparallel.
Once relativistic effects are included and/or the ferromagnetic state is 
considered, this simplicity is lost and the continuous probability distribution
$P^{(\hat{n})} (\hat{e}_i)$'s must be sampled for a fine mesh of angles 
and the averages with the probability distribution performed numerically.
(Careful checks have to be made to ensure that the sampling of 
$P^{(\hat{n})} (\hat{e}_i)$ is sufficient - in our calculations
some 25,000 values were used.)
In the ferromagnetic state, the magnetic anisotropy is given by the 
difference between the free energies, $F^{(\hat{n})}$,
for different magnetisation directions, $\hat{n}$, but the same 
magnetisation $M$.

Once again our study of $FePt$ starts with a self-consistent,
scalar-relativistic calculation, this time for the paramagnetic (DLM) state. 
On the $Fe$ sites a local moment of 2.97 $\mu_B$ is set up whilst no moment 
forms on the $Pt$
sites. For the same lattice spacings ($c=0.385$nm, $c/a =1$), we found that, 
for the completely ferromagnetically ordered state of $FePt$ at $T=0$K, the 
magnetisation per $Fe$ site is 2.93 $\mu_B$ and a small magnetisation of 0.29
$\mu_B$ is associated with the $Pt$ sites. This suggests that the thermal
effects on the magnetic properties should be well-described by the DLM picture.
Using the self-consistent potentials and magnetic fields of the paramagnetic DLM
state, we proceed by picking a
series of values of $M(T)/M(0)$ to set the probabilities, 
$P^{(\hat{n})} (\hat{e}_i)$. By calculating $\langle \Omega^{(\hat{n})}
\{ \hat{e}_l \} \rangle_{\hat{e}_i}$, using our relativistic DLM method, 
we find the Weiss field, $h^{\hat{n}}$, at each value and for a given $\hat{n}$,
and hence determine the temperature dependence of the magnetisation.
The results are shown in Figure 1. Although the
shortcomings of the mean field approach do not produce the spinwave $T^{
\frac{3}{2}}$ behavior at low temperatures, the easy axis for the onset of
magnetic order is obtained ( $h^{(001)} > h^{(100)}$ as $T \rightarrow T_c$)
and it corresponds to that found at lower temperatures
both experimentally and in all theoretical ($T=0$K) calculations, 
perpendicular to the $Fe$, $Pt$ layer stacking. (A straightforward adaptation
of this approach to other systems such as thin films in combination with $T=0$K
calculations may be useful in understanding temperature induced spin 
reorientation transitions.) A Curie temperature
of 935K is found, in fair agreement with the experimental value of 750K.
~\cite{DLM,Onsager}

\begin{figure}[tbh]
\resizebox{0.9\columnwidth}{!}{\includegraphics*{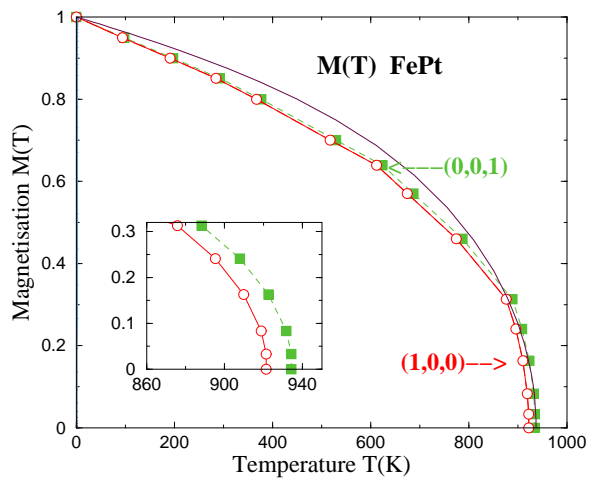}}
\caption{The magnetisation of ordered FePt as a function of temperature. The 
filled squares refer to a magnetisation along $\hat{n}= (0,0,1)$ whilst the
open circles refer to $\hat{n}= (1,0,0)$. The onset of magnetic order 
occurs at 935K along the easy axis, $(0,0,1)$. The full line shows the 
magnetisation from the mean field approximation to a classical Heisenberg 
model for comparison. The inset shows the region near the onset.}
\end{figure}

The free energy difference, $F^{(0,0,1)}-F^{(1,0,0)}$, i.e. the 
magnetocrystalline anisotropy, $K(T)$, is also calculated using the
theory and leads to the key results of this letter shown in Figure 2. At $T=0$K,
the MCA has a value -1.740 meV, close to the value, -1.696 meV,
obtained by the earlier, separate calculation for this situation
for the completely ferromagnetic state. Figure 2
shows the MCA, $K(T)$, versus $M(T)/M(0)$. Curves for the single-ion 
classical spin model anisotropy~\cite{Callen} and $(M(T)/M(0))^{2}$ are also
shown for comparison. Apart from near complete magnetisation,
$0.9<M(T)/M(0) < 1$,  our results 
show a good approximate $(M(T)/M(0))^2$ behavior in good agreement with
experiment.~\cite{Okamoto,Thiele2002}. This is in marked contrast 
to the model, which becomes proportional to  
$(M(T)/M(0))^{3}$ for the larger $M(T)$'s. 
  
\begin{figure}[tbh]
\resizebox{0.9\columnwidth}{!}{\includegraphics*{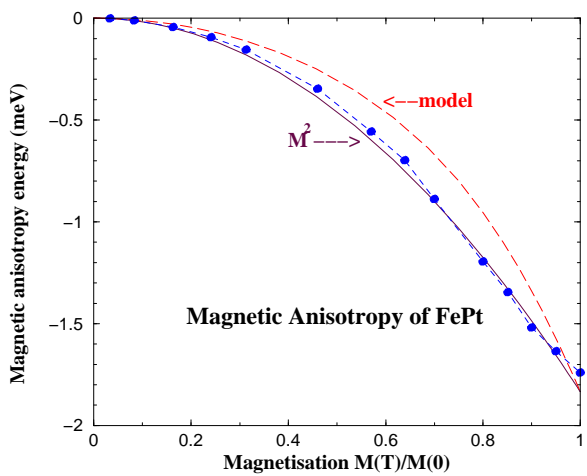}}
\caption{The magnetic anisotropy of FePt as a function of magnetisation. The 
filled circles show the calculations from the ab-initio theory. The full line
show $K_0 (\frac{M(T)}{M(0)})^2$ and the dashed line shows the single-ion model
function $K_0 \frac{< g_2 (\hat{e})>_T}{ < g_2 (\hat{e})>_0}$ with $K_0=$ -1.835 meV.}
\end{figure}

Evidently Figure 2 shows that at low temperatures 
in the single ion model the MCA falls off much more 
quickly as the temperature is increased and the overall
magnetisation is reduced.
 Moreover our theory for an itinerant electron
system does capture the behavior of the $K$ versus magnetisation 
relation quantitatively. This
theory assumes that there is a separation between fast and slow electronic
degrees of freedom. A picture of `local moments' emerges naturally but with a 
subtlety that their existence and behavior are determined by the fast
electronic motions. We expect the MCA of the important magnetic 
materials $L1_0$-CoPt and FePd to follow a similar variation with magnetisation
since the local moments sustained on the Co and Fe sites in the paramagnetic
DLM states (1.78 and 2.98 $\mu_B$ respectively) are comparable in size to 
magnetisation per site in the completely ferromagnetic states (1.91 and 
2.96 $\mu_B$). Our DLM theory should therefore have good prospects in 
describing the variation of $K$ with magnetisation for a range of metallic
magnets like these.
The success of the above relativistic DLM methodology in 
explaining the unexpected behavior of $L1_0$-
$FePt$ suggests that further calculations for promising 
magnetic materials in bulk, thinfilms or in magnetic 
nanostructures may be valuable for the future modelling and 
exploitation of their magnetic properties.

We acknowledge support from the EPSRC(U.K), CSAR, the Centre
for Scientific Computing at the University of Warwick and the
Hungarian National Science Foundation (OKTA T046267).

\end{document}